\pdfoutput=1 
\documentclass{JINST}
\usepackage{fontenc}

\usepackage[utf8]{inputenc}
\usepackage{graphicx}

\usepackage{amssymb}

\usepackage[figuresright]{rotating}

 \title{Tests of PMT Signal Read-out
of Liquid Argon Scintillation with a New Fast Waveform
Digitizer }

\author{R. Acciarri$^a$, N. Canci$^b$, F. Cavanna${^a,}{^c}$, A. Cortopassi$^d$, M.~D'Incecco$^b$, G. Mini$^d$, F.~Pietropaolo$^e$, A. Romboli$^d$, E.~Segreto$^b$, A.M. Szelc$^c$\thanks{Corresponding Author}~\thanks{Previously at IFJ PAN  and Universit\`a dell'Aquila} \\
\llap{$^a$} Universit\`a dell'Aquila and INFN, \\
  L'Aquila, Italy,\\
\llap{$^b$} INFN - Laboratori Nazionali del Gran Sasso, \\
  Assergi, Italy,\\
\llap{$^c$} Yale University, \\
   New Haven, CT\\
\llap{$^d$} CAEN SpA \\
  Viareggio, Italy \\
\llap{$^e$} INFN - Sezione di Padova,  \\
  Padova, Italy\\

  E-mail: \email{andrzej.szelc@yale.edu}}

\abstract{
The CAEN V1751  is a new generation of Waveform Digitizer recently introduced by CAEN SpA.  
It features 8 Channels per board, 10~bit, 1~GS/s using Flash ADCs Waveform Digitizers 
(or 4 channels at 2~GS/s in Dual Edge 
Sampling mode)  with threshold and Auto-Trigger capabilities. This
 provides a good basis for data acquisition in Dark Matter searches using 
PMTs to detect scintillation light in liquid argon, as it matches the requirements for measuring the fast scintillation component. The board was tested by operating it in real experimental conditions and by comparing it with a state of
the art digital oscilloscope. We find that the sampling at 1 or 2~GS/s is appropriate for the reconstruction of the fast 
component of the scintillation light in argon (characteristic 
time of about 6-7 ns) and the extended dynamic range, after a small customization,
allows for the detection of signals in the range of energy needed. The
bandwidth is found to be adequate and the intrinsic noise is very low.
}


\begin{document}

\pdfinfo{%
  /Title    (Tests of PMT Signal Read-out
of Liquid Argon Scintillation with a New Fast Waveform
Digitizer)
  /Creator  ()
  /Producer ()
  /Subject  ()
  /Keywords ()
}




\section{Introduction}

Liquid argon (LAr) is currently one of the quickest developing technologies in both Dark Matter and neutrino detectors. This, particularly in the case of Dark Matter applications, can be ascribed to the possibility of detecting both the ionization and scintillation light resulting from particle interactions. Both of these signatures can give a strong handle on background discrimination and registering the scintillation light in particular can provide a lot of information and give another method to suppress the background, especially at lowest energies which are crucial in Dark Matter searches. However achieving and understanding these lowest energies make preparing the data acquisition setup for liquid argon detectors a very challenging task.

The first challenge comes from the fact that the PMTs or any light collection elements must be suited to operate in cryogenic temperatures, as low as the liquid argon boiling temperature of 87 K. Once this is accomplished, the next challenge is having a fast enough acquisition system because of the light emission mechanism in liquid argon, where the scintillation light is a result of deexcitations of argon dimers. There are two main molecular states: the singlet $^1\Sigma_u$ and the triplet $^3\Sigma_u$ \cite{hitachi}. These two states have radically different 
decay times - 5$\div$7 $ns$ and 1.2$\div$1.5~$\mu
s$, respectively, which results in a two component form of the scintillation light. This fact is particularly interesting to Dark Matter applications, because the ratio of the relative amplitudes of these states is
usually different for different types of particles and so it is often considered useful for Pulse Shape Discrimination. 
 Another requirement or challenge for a prospective DAQ system would be the ability to identify single photoelectrons (p.e.), since at lowest energies they would form a
significant part of the event. 
For these reasons most of the liquid argon Dark Matter experiments
have chosen to employ fast waveform digitizers of up to 1~GS/s sampling
\cite{WArP},\cite{clean} and \cite{ArDM}.

In this work we report tests performed on a digitizer with 1~GS/s sampling - the new
V1751 fast ADC board developed by CAEN SpA. The board has been operated in working conditions
in a liquid argon Dark Matter test detector. In order to test if the sampling implemented
in the board is sufficient to correctly reconstruct single photoelectrons it has also been
compared to a state of the art oscilloscope capable of much higher sampling
rates - up to 20~GS/s. 

This work first describes the new board and the results of tests performed when using the
board in an acquisition chain in a liquid argon detector for low energy PMT signal readout. The
data obtained using the V1751 board are then compared to runs taken with a LeCroy
WavePro 735Zi oscilloscope to check whether the bandwidth and sampling provided are
sufficient to acquire single photoelectrons in liquid argon.

\section{The V1751 Fast Waveform Digitizer}



\begin{figure}[ht]
 \centering
 \includegraphics[height=10cm]{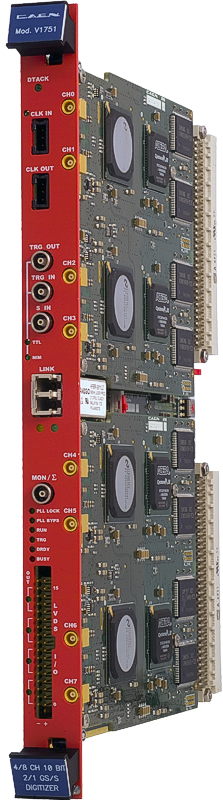}
 \includegraphics[height=10cm]{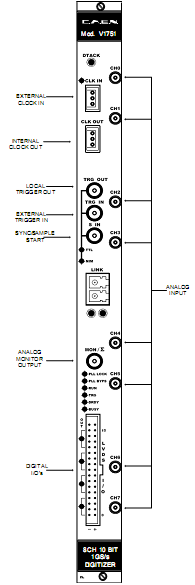}
 \caption{CAEN V1751 4/8 Channel 10 bit 2/1~GS/s Digitizer and its front panel.}
\label{fig:board}
\end{figure}

\begin{figure}[ht]
 \centering
 \includegraphics[width=10cm]{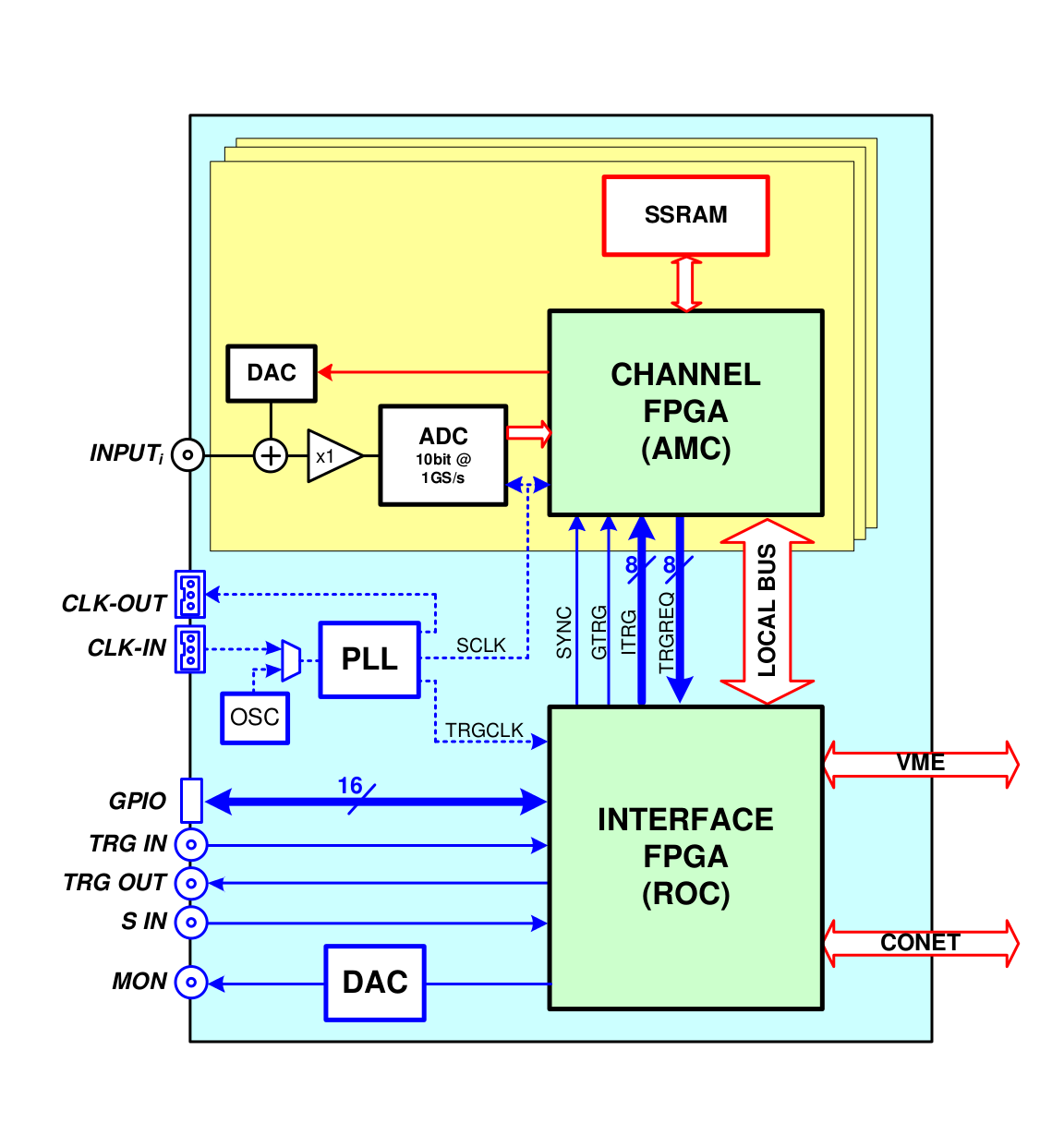}
 \caption{CAEN Digitizer block diagram. The mother board defines the
form-factor; it contains one FPGA for the readout interfaces and the
services.The daughter board defines the type of digitizer; it contains the input
amplifiers, the ADCs, the FPGA for the data processing and the memories}
\label{fig:bdesign}
\end{figure}

The CAEN V1751 is a 1 unit wide, 6U VME  board housing a 8 channel 10~bit 1~GS/s Flash ADC waveform digitizer, see figure \ref{fig:board} with an analog bandwidth of 500 MHz and an input
dynamic range of 1 V$_{pp}$. The single ended version of the board allows adjusting
the DC offset of each channel by means of an internal 16-bit DAC in the range
 -0.5/+0.5~V.
By interleaving pairs of channels, the board can run in Dual Edge Sampling
(DES) mode: in this way, the board works as a 4 channel 10~bit 2~GS/s
digitizer.
The V1751 has been designed with a mother-daughter board configuration - see
figure \ref{fig:bdesign}. The mother board is equipped with an FPGA
dedicated to the readout interfaces and services such as power supplies, clocks
and I/Os. 
Each of the daughter boards houses 4 channels and contains the input
amplifiers, 
the ADCs and the FPGAs for data processing. Each channel has a SSRAM
memory buffer (1.8 or 14.4 MS/ch depending on chip installed on the board), with
independent read-write access divided into 1 to 1024 buffers of programmable size. 
$1.8 MS/s$ corresponds to $\simeq 370 cm $ of drift length at 1 kV/cm. Thanks to the buffers, this memory can be assigned to one or many waveforms, in the second case miminizing the dead time if multiple events occur too close in time to read them out. It can also give time for the onboard FPGA to perform compression on the data. As an example, with the 1.8 MS/ch chip the memory could be divided into 5 segments of 900kS/s, sufficient for most next generation Dark Matter detector designs wanting to register secondary scintillation. In case this would not be sufficient the board could be fitted with the 14.4 MS/ch chip effectively increasing the memory by a factor of 8.  In DES mode the memory buffers of each couple of interleaved channels are
shared de facto doubling the memory depth per channel. 

The trigger signal can be provided via the front panel External TRG IN input as
well as via the VMEbus. It can also be generated internally, allowing an independent
self-generation of a majority trigger.
Being designed with multi-board synchronization capabilities, the trigger from
one board can be propagated to the other boards through the front panel Trigger
Output; in this way, it is possible to compose a complete DAQ system with a
large number of channels.

A programmable Analog Output allows to reproduce the trigger majority, a test
waveform and the buffer occupancy. 
The module’s VME interface is VME64X compliant. 
The board houses a daisy chainable Optical Link able to transfer data
up to 100 MB/s through a CAEN proprietary protocol\footnote{in the measurements reported here a board with an older version of the firmware using the CONET protocol was used.} called CONET 2 , thus it is possible to
connect up to eight ADC boards (64 ADC channels) per link by means of
CONET-compliant optical controllers, (CAEN Mod. A2818 or Mod. A3818). 
Optical Link and VME access are internally arbitrated. 

\begin{figure}[ht]
 \centering
 \includegraphics[width=10cm]{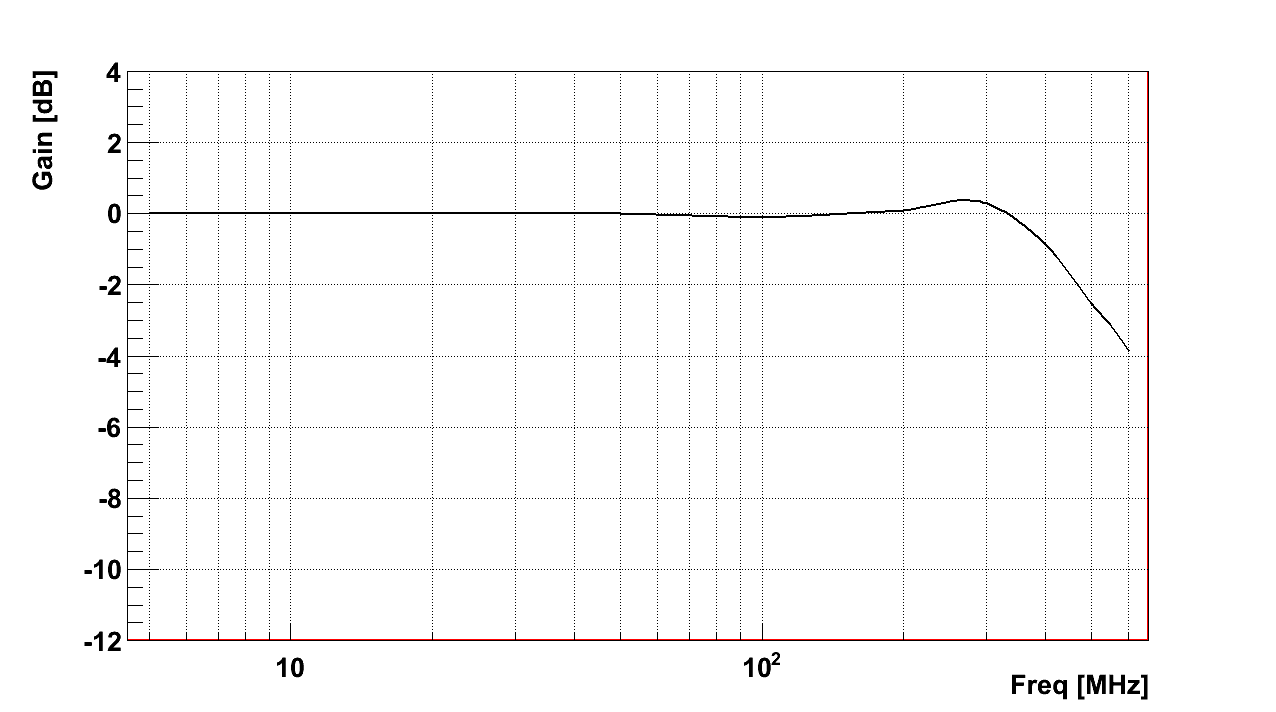}
 \caption{Full Power Bandwidth graph of the modified front end stage. The 500
MHz bandwidth is preserved.} \label{fig.band}
\end{figure}

In order to adjust the dynamic range of the board with the PMT output in a way to be able to see single photoelectrons at the standard PMT gain used, a
customization of the front end stage of the board has been required. 
Modifying the input amplifiers, the dynamic range has been reduced to 0.2 V$_{pp}$ from the standard 1 V$_{pp}$
preserving the analogue bandwidth, see figure \ref{fig.band}. The smaller dynamic range should still be sufficient to observe secondary signals, even though they provide much more light than primary signals. This is because the secondary scintillation light arrives in a much broader time window than the primary scintillation and so, since it is not integrated, it is unlikely that it would saturate the ADC range for signals where the primary signals do not do so as well. This is even less likely for larger detectors since there the light should be spread out among a larger number of photodetectors.

\section{Tests in Experimental Conditions}

In a Dark Matter search it is important to be able to register the lowest energy events and, in the case of liquid argon, to be able to understand the fast pulses coming from PMTs registering the fast component of the scintillation light. 
This means that a DAQ board should have parameters that are tailored to this task. Specifically, the sampling and noise level of the board are important in order to observe single photoelectron pulses correctly and not lose them
amongst the noise counts. A sufficiently big full scale range is also useful in order to
observe higher energy events - which becomes imporant in high light yield (LY) per PMT setups \cite{PMT_paper}.
WIMP-like or neutron-like recoil event collection will be affected even more since the scintillation light in these events is dominated by
the fast singlet component and so a large part of the light arrives almost
simultaneously with the trigger which makes saturation more probable.


\subsection{The Experimental Setup}

\begin{figure}[ht]
 \centering
 \includegraphics[width=10cm]{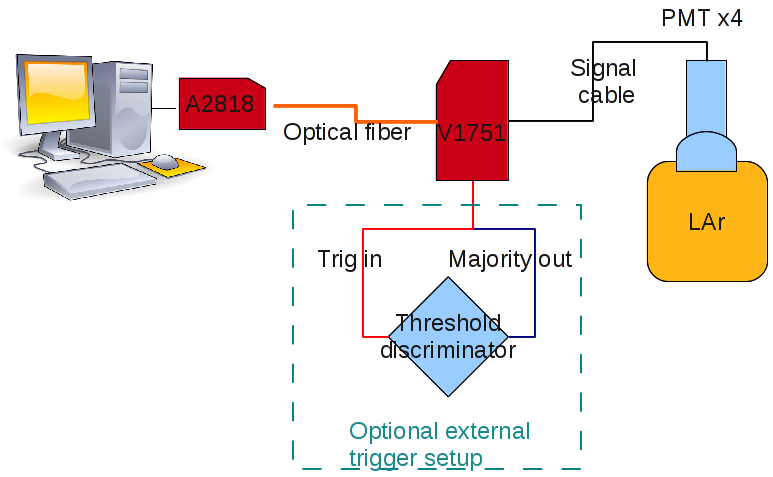}
 \caption{The experimental setup used in these measurements.}
\label{fig:exp_setup}
\end{figure}

To our knowledge the setup used in this work, see figure \ref{fig:exp_setup}, 
 is the first use of the V1751 board in working conditions analogous to those of a Dark Matter detector. A liquid argon volume of 4.3 liters was instrumented with four high Quantum Efficiency (QE) Hamamatsu manufactured R11065 PMTs with a negative
bias on the cathode. The anode signals were fed directly into four of the V1751 input channels. A trigger occured when a majority of 3 out the 4 channels registered a signal of the order of 1 photoelectron at a PMT gain of $0.8 \cdot 10^6$ (pulses as low as $\simeq$ 2 mV). 
The trigger condition was first asserted by
feeding the majority output generated by the V1751 board into a threshold
discriminator and feeding back the resulting NIM signal into the TRG IN input of the
board. In the second phase of the measurement the majority testing was performed
internally inside the ADC board by polling a board register address. The
acquired events were then transferred via optical fibre into an A2818 CAEN PCI board
connected to a PC computer.

The 10 bit Full Scale (FS) of the board allowed observing events in a range from 20 p.e. to 6000 p.e. for gamma like
events. The lower limit is actually attributed to the inefficiency of the reconstruction software. The board trigger is capable of triggering on pulses with a height of the order of single p.e.. in a 4~PMT setup corresponding to a 3.3~$\div$~1000 keV$_{ee}$ at light yield equal
to 6.1 \cite{PMT_paper} (20~$\div$~1000 p.e. for recoil like events corresponding to 13~$\div$~655 keV nuclear) energy range - sufficient for a Dark Matter
detector.

\subsection{Data Quality Tests}

\begin{figure}[ht]
 \centering
 \includegraphics[width=6.5cm,height=4cm]{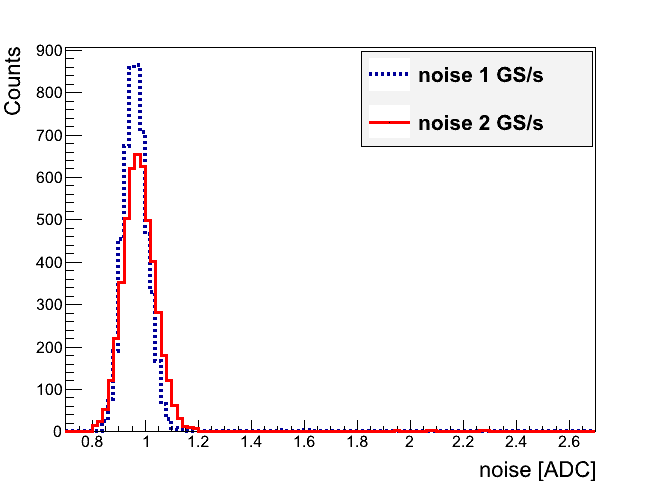}
 \includegraphics[width=6.5cm,height=4cm]{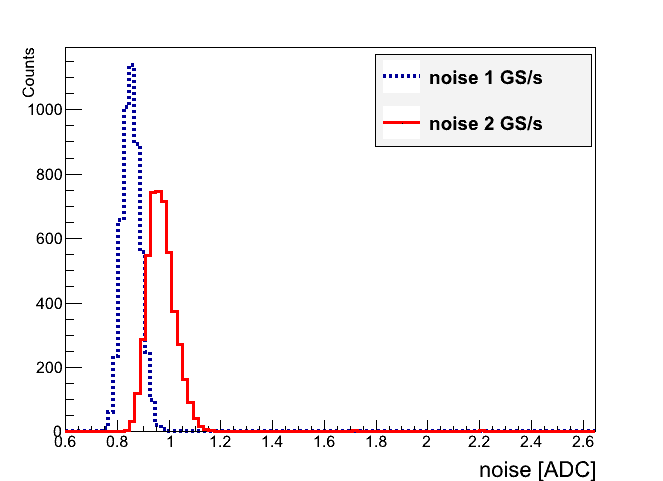}
 \caption{The noise, defined as the RMS of the baseline, measured in the board for two different channels for the 1~GS/s [blue-dashed] and 2~GS/s [red] mode of the board.}
\label{fig:noise_channel}
\end{figure}

The first tests were concentrated on the basic quality of the
acquired data. One of the quantities measured was the noise defined as the RMS of the
baseline in the pretrigger portion of the recorded waveforms, set at 2~$\mu s$. It has been
found to be under 1 ADC i.e. under the LSB (Least Significant Bit) as declared
by the manufacturer. The values were slightly worse when running in the 2~GS/s
DES mode, see figure \ref{fig:noise_channel}. 

\begin{figure}[ht]
 \centering
\includegraphics[width=6.5cm]{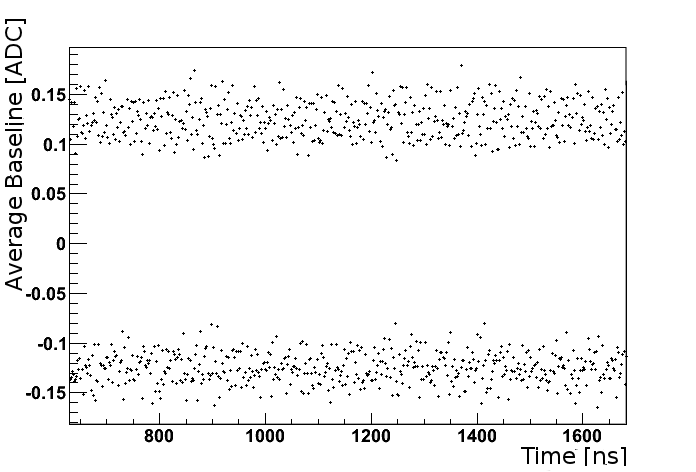}
\includegraphics[width=6.5cm]{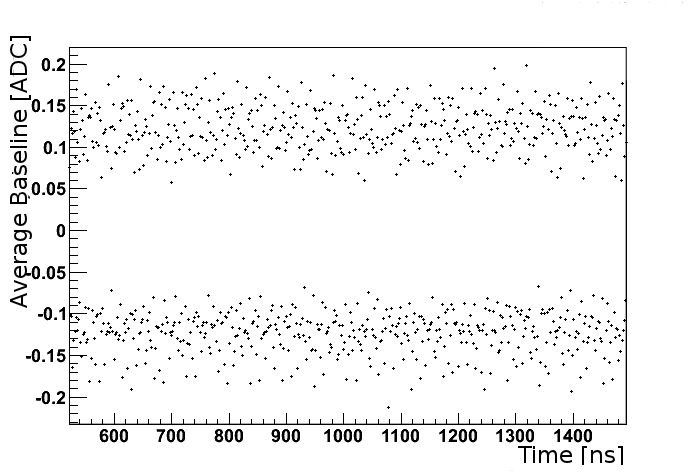}
 \caption{The dephasing effect observed
when summing multiple events together for a single channel [left] and four channels combined [right]. }
\label{fig:dephasing}
\end{figure}

In a subsequent test consecutive events were summed up to create an average waveform, which can be 
used to estimate the purity of liqid argon \cite{N_2,O_2}. 
This operation allowed observing a dephasing effect in the
baseline region, where the odd and even timesample values seemed to be separated by an
offset even when summing the 4 channels together, see figure
\ref{fig:dephasing}. The reason for this behaviour is attributed to the way the board
sends out the acquired samples in pairs. It should be noted that the
observed offset in the average waveform is smaller than 1 ADC per event and so
should not cause problems on an event per event basis and can be easily corrected for
by software means. 


The small cross-sections for Dark Matter interactions lead to a scaling up of detector mass 
 which for liquid argon \cite{ArDM},\cite{DarkSide} usually translates to longer electron drift times in the detector. 
This means that longer waveforms will need to be acquired, therefore we have tested the data throughput obtainable with the board. 
We have found that acquiring with an automatic software trigger results in a 17~$Hz$
acquisition rate for 8 channels of 400~$\mu s$ waveforms corresponding to a ~49.6~$MB/s$ data
rate. This is less than the declared throughput of the CONET protocol used by
the board to communicate with the PC. However the
limit was a result of the speed of writing to disk. Since the event rate in a liquid argon detector is dominated by a large background rate of about 1~$Bq/kg$ coming from the radioactive isotope of
$^{39}$Ar \cite{Ar39} future detectors will probably need to implement a second level trigger, data decimation or compression or use depleted argon~\cite{Depleted_ar} in order to sustain the data throughput.


\begin{figure}[ht]
 \centering
 \includegraphics[width=6cm]{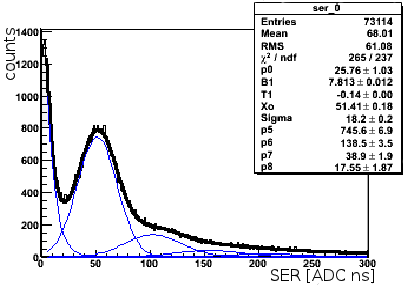}
\includegraphics[width=6cm]{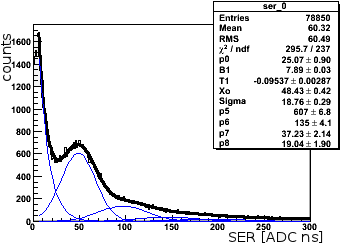}
 \caption{SER spectrum obtained with the board [left] compared to the old DAQ setup [right] }
\label{fig:ser}
\end{figure}


\begin{figure}[ht]
 \centering
 \includegraphics[width=9.0cm]{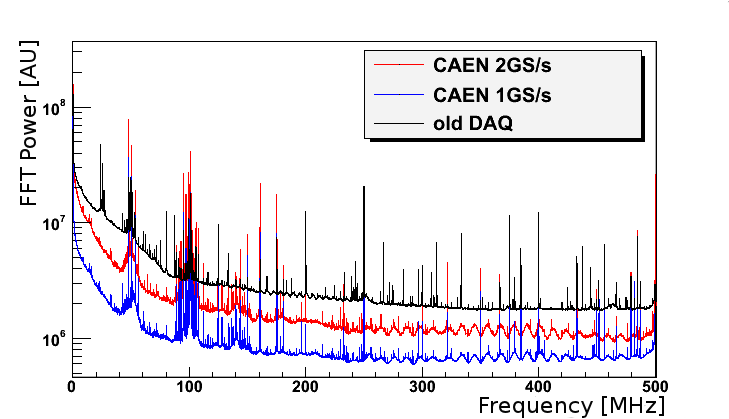}
 \caption{The frequencies of
the noise found in the board at 1~GS/s [blue - lowest], 2~GS/s DES mode [red - middle] and
previously used setup [black - highest] obtained by performing an FFT of no-trigger events. The forest of lines around 100MHz is attributed to noise from commercial radio stations.}
\label{fig:FFT}
\end{figure}

The single electron response spectra (SER), which are a standard diagnostics tool for PMT based DAQ setups, obtained with the board were clear with a peak well separated from the steep exponential dark counts - figure \ref{fig:ser}.  The new
board showed an improvement with regard to the previous generation DAQ setup used \cite{PMT_paper}. This is
attributed to the much better noise suppression in the V1751 board setup
confirmed by an FFT transform on waveforms acquired without trigger, therefore containing mostly ambient noise - figure \ref{fig:FFT}.

\section{Liquid Argon Signal Readout at Higher Samplings}

To fully test the performance of the V1751 board  we decided to
compare it to a state-of-the-art digital oscilloscope to see if the bandwidth and sampling
rate of the V1751 board are really sufficient to register liquid argon signals. A
LeCroy WavePro~735Zi oscilloscope was chosen as the reference. It is
capable of achieving 20~$GS/s$ sampling
(50~$ps/sample$) simultaneously with 4 channels at a 3.5~$GHz$ analog bandwidth and 8 bit Full Scale.
However, due to its price and operation mode it is not conceived for use in an actual data
acquisition.

As a first step we compared the value of the single photoelectron integral
obtained by fitting the SER spectrum resulting from binning the single peaks
found in the examined waveforms. 

To perform this measurement the PMTs were set to a higher gain than in the usual data taking - $2 \times 10^7$
to focus on the single photoelectron signals. Additionally the waveforms were acquired in a time period between 5 $\mu s$ and 7 $\mu s$ after the trigger to focus on the tail of the slow component of the scintillation light in liquid argon (with an exponential decay time of 1.2~$\mu s$ \cite{N_2}) and to save disk space. The obtained waveforms were
scanned using a simple peak finding algorithm and the integrals of the peaks
found were binned into a spectrum after local baseline subtraction, which was estimated in a 50 ns region. If subsequent peaks were closer together than 50~$ns$ they were merged and so the resulting spectrum was fit with a superposition of multiple (N) gaussian peaks
corresponding to the first and multple p.e. peaks and an exponential
function to account for the component due to dark counts and baseline fluctuations. The position and variance of
the n$^{th}$ peak was constrained to the values resulting from the single p.e. peak
as in $x_n = n x_0$ and $\sigma_n = \sqrt{n} \sigma_0$. 

\begin{figure}[ht]
 \centering
\includegraphics[width=9cm]{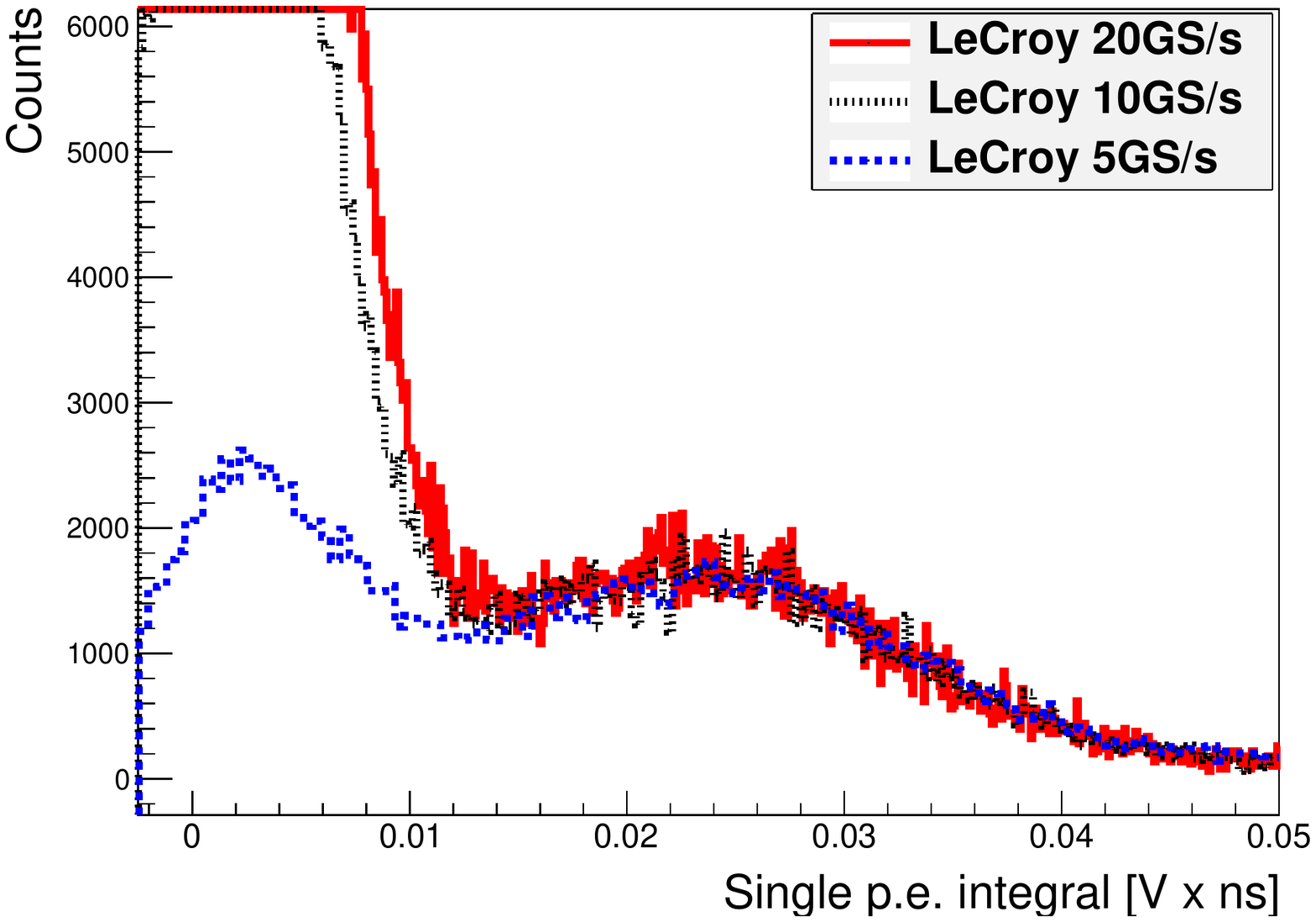}
 \includegraphics[width=9cm]{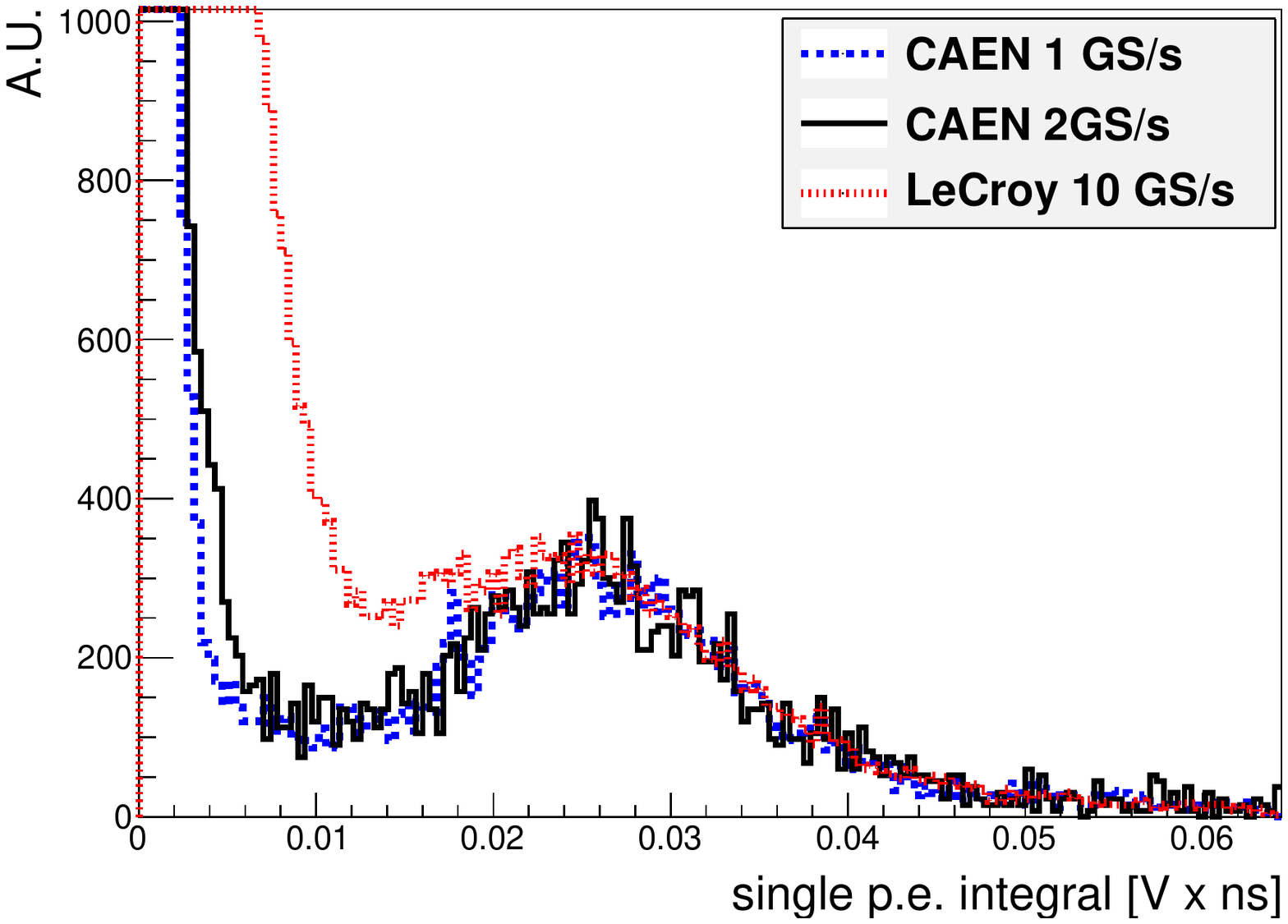}
 \caption{Comparison of the Single Electron Response (SER) spectrum position for
different sampling
frequencies: 20~GS/s [red-solid], 10~GS/s [black-dotted] and 5~GS/s [blue-dashed] taken with the LeCroy
oscilloscope [left] and the 10~GS/s SER [red-dotted] spectrum compared with the waveforms
obtained using the V1751 board at 2~GS/s [black-solid] and 1~GS/s [blue-dashed]. The CAEN board SER plots have been rescaled to $V \cdot ns$ units of the Oscilloscope. }
\label{fig:SERs}
\end{figure}

The resulting single p.e.
peak positions obtained at high sampling frequencies (5~GS/s, 10~GS/s and 20~GS/s) with the oscilloscope are in good agreement with each other, see figure \ref{fig:SERs} [left], proving that the sampling rate should not affect the SER peak determination. The SER positions measured with the V1751 board, after converting to $V \cdot ns$, differed by at most 10\%, a relatively good agreement, with those obtained using the oscilloscope suggesting that the lower sampling rate of the V1751 board does not affect the
 PMT charge collection capabilities of a potential DAQ setup. The difference of the absolute position should not affect the actual energy measurements of the detector, since event waveforms are normalized to the SER position. The higher bandwidth of the LeCroy oscilloscope actually increased the fraction of noise peaks selected by the algorithm, due to high frequency noise making the single p.e.
position determination more difficult - see figure \ref{fig:SERs} [right], where the noise shoulder is much more pronounced in the LeCroy data.

\begin{figure}[ht]
 \centering
 \includegraphics[width=9cm]{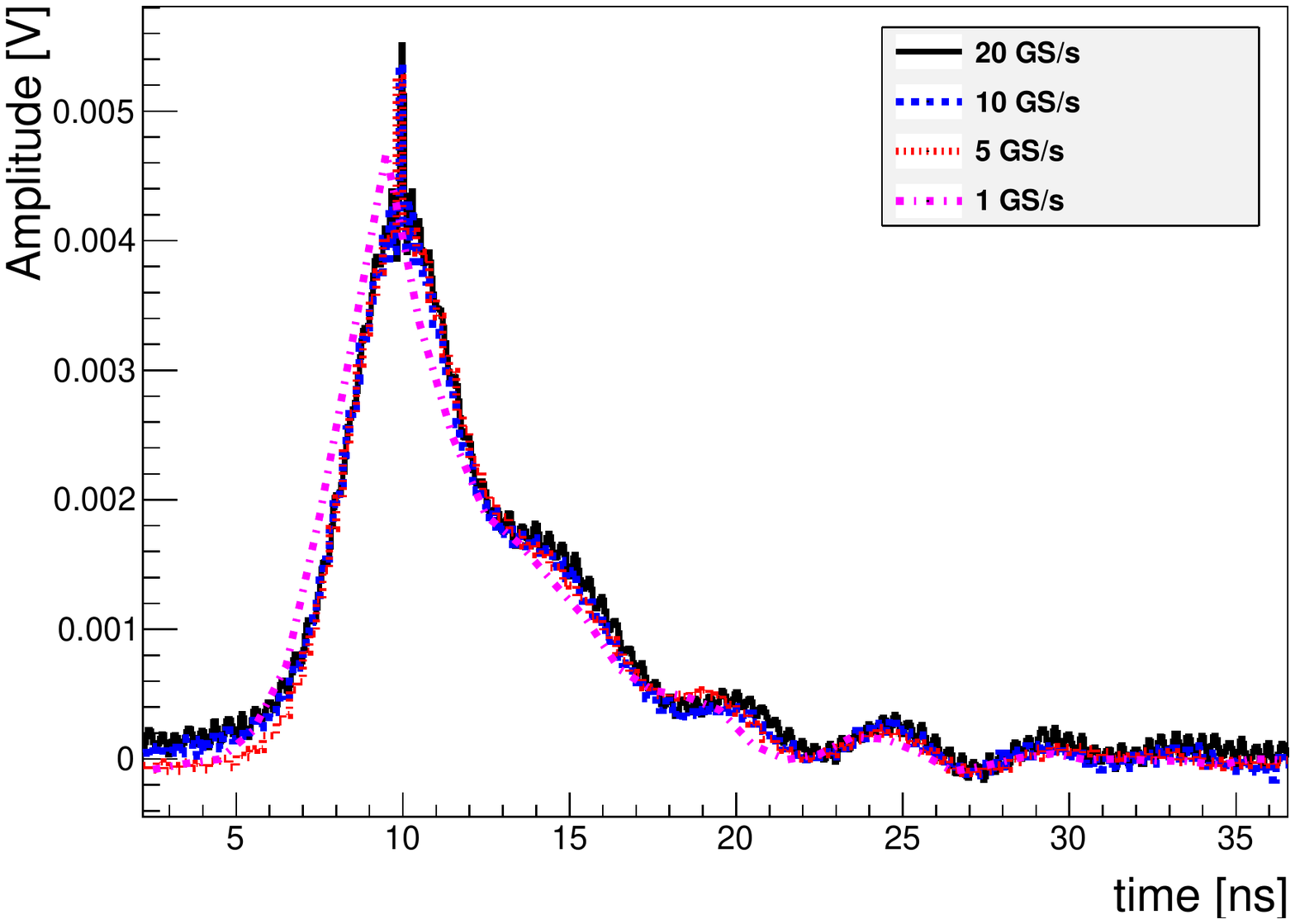}
\includegraphics[width=9cm]{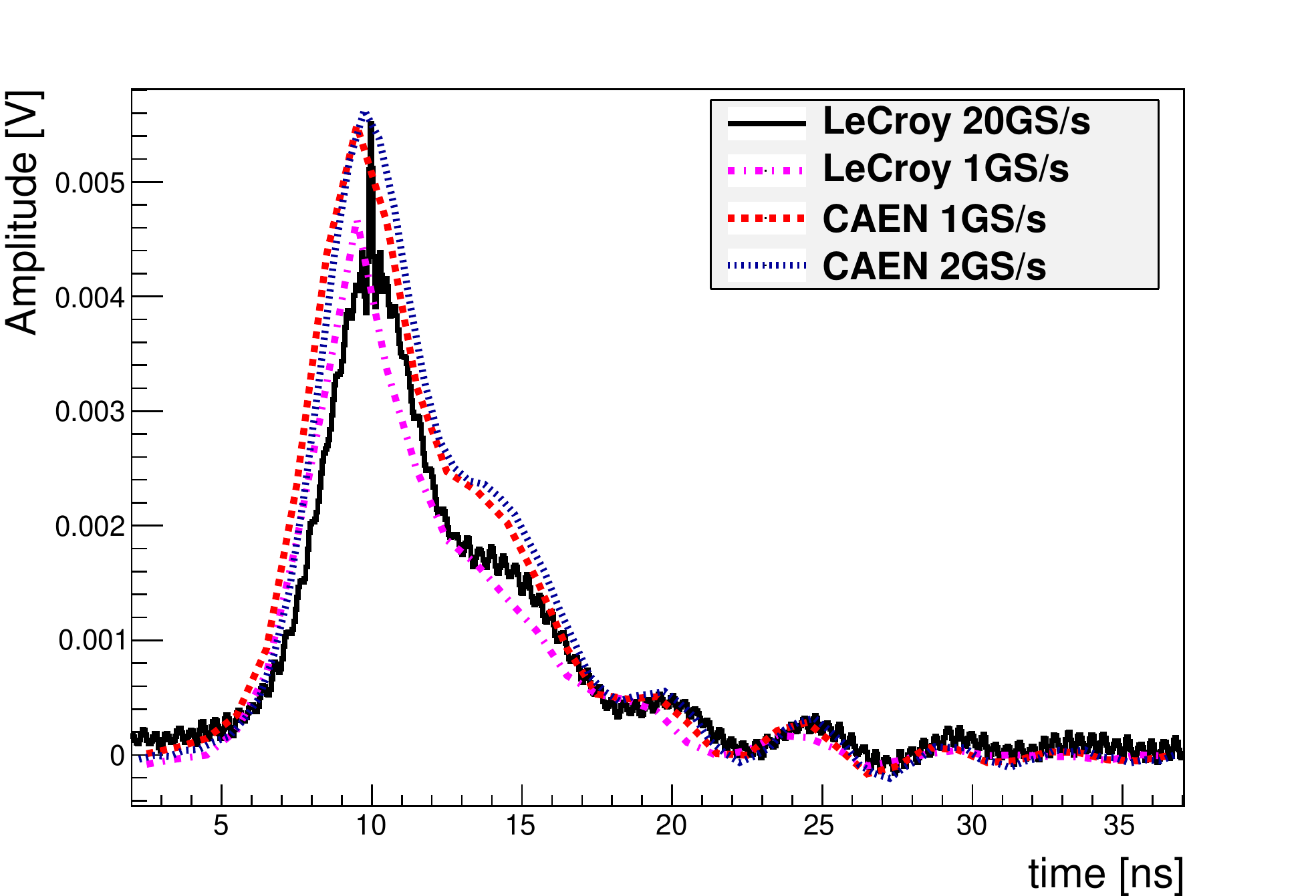}
 \caption{Comparison of the average single p.e. shape for different sampling
frequencies: 20~GS/s [black-solid] , 10~GS/s [blue-dashed] , 5~GS/s [red-dotted] and 1~GS/s [magenta-dotted-dashed] taken with the LeCroy
oscilloscope [left] the average 20~GS/s [black-solid] and 1GS/s p.e. [magenta-dotted-dashed] waveforms compared with the waveforms
obtained using the V1751 board at 2~GS/s [blue-dotted] and 1~GS/s [red-dashed [right]. }
\label{fig:single_phel_wvfm}
\end{figure}

\begin{figure}[ht]
 \centering
 \includegraphics[width=8cm]{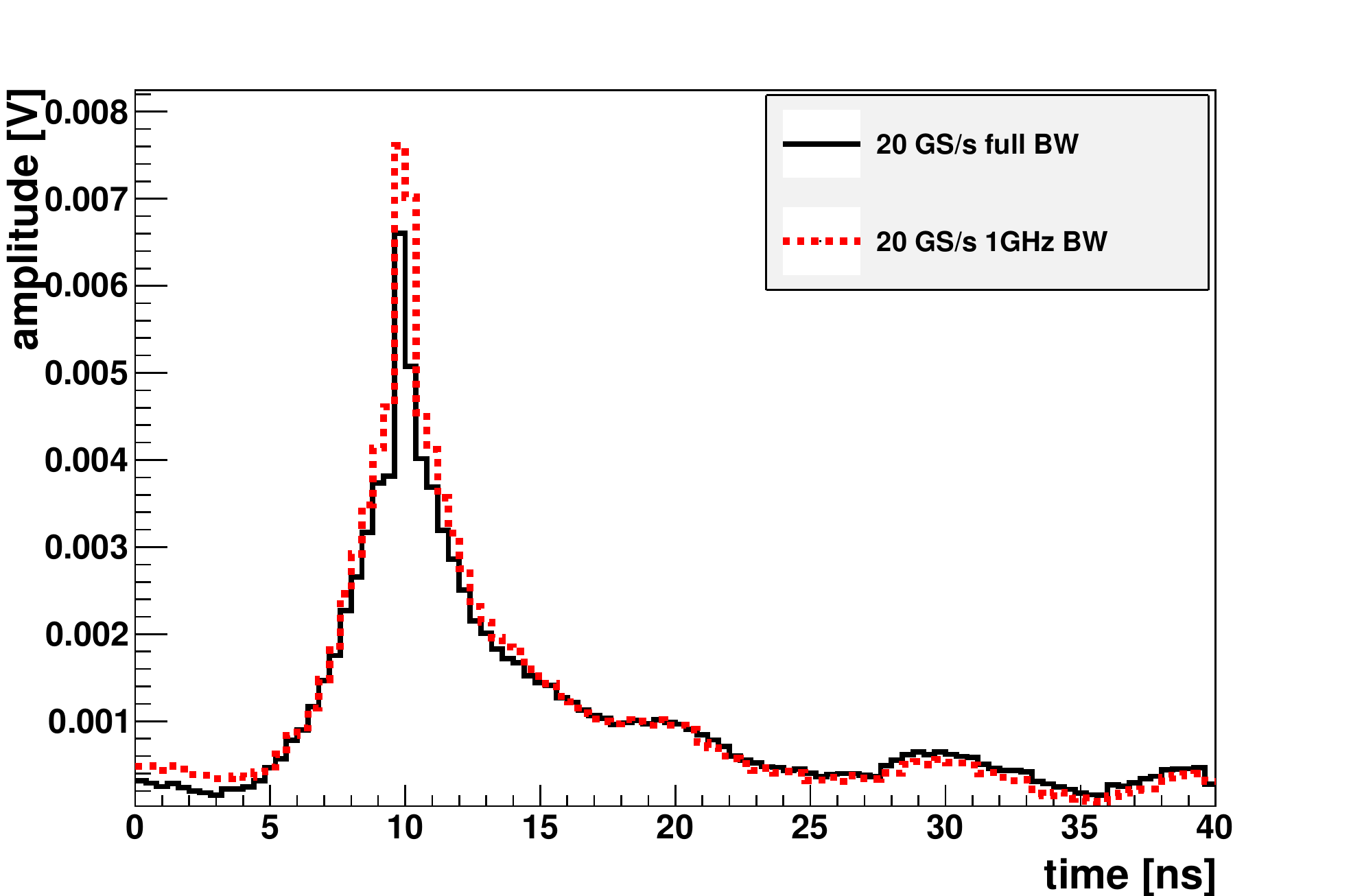}
 \caption{ A comparison of 20~GS/s average pulses, rebinned to 1~GS/s for clarity, obtained with the LeCroy oscilloscope at 3.5 GHz bandwidth - [black-solid] and 1 GHz bandwidth - [red-dashed]. }
\label{fig:bandwidth}
\end{figure}

The next performed test had the objective of determining whether with 1~GS/s sampling some
detail in the shape of the single p.e. signal is not lost. 
Using the single photoelectron peak parameters obtained from the SER spectrum evaluation the waveforms
were rescanned to search for peaks above the noise level and with an integral in
the range of ($x_o-\sigma$ ; $x_o+\sigma$) resulting from the fit of the first
photoelectron peak. In this analysis, care was taken to discard peak pairs closer than 100~$ns$.  
For the V1751 data this operation was performed on a sample
of 100 000 offtrigger 15 $\mu s$ long waveforms. For the LeCroy data, the waveforms used were the same as for the SER determination at samplings of 20~GS/s, 10~GS/s, 5~GS/s and
1~GS/s. Five thousand waveforms were saved for each PMT and sampling. The resulting peak waveforms were then summed together and divided by the number of events to obtain an average single photoelectron pulse in V units. 
The waveforms obtained with the oscilloscope can be seen in figure
\ref{fig:single_phel_wvfm} [left], where the agreement between the
different
sampling times is quite satisfactory. In figure \ref{fig:single_phel_wvfm} [right], the 20~GS/s single p.e. waveform is compared 
to those obtained using the V1751 in standard and DES mode. In these units the V1751 pulse is slightly higher, which is attributed to the lower
bandwidth of the V1751 board which results in less fast frequency noise that
enters into the waveform diluting its amplitude. This is consistent with the higher values of the SER peak obtained with the V1751 board as in figure \ref{fig:SERs}.  This effect has been
confirmed by looking at single phe waveforms obtained with the LeCroy
oscilloscope setting different bandwidth filter settings, see figure \ref{fig:bandwidth}.


Light yield is an important parameter in a
scintillation detector because it expresses its light detection efficiency with
respect to the energy deposited in the chamber. It is measured in units of
p.e./keV usually by irradiating a chamber with a radioactive
source and comparing the detector response to the source energy.
 For this test a $^{241}Am$ source with a monoergetic gamma peak at 59.5~$
keV$ was used. At this energy the emitted $\gamma$-rays undergo full photoelectric
conversion depositing all of their energy in the liquid argon chamber. For
details of this calibration we refer to \cite{PMT_paper}. A high LY is extremely important  in a Dark Matter
 detector since the higher 
it is the lower the incident particle energies a detector can register and
identify. The LY is mostly affected by detector components i.e. the PMT or
wall reflection efficiency, but one can imagine that if the electronics setup
were to miss a fraction of photoelectrons due to e.g. their shape or size - the
effective LY could be diminished. For this reason, the last test of the new setup was to check 
whether the 1~GS/s sampling of the V1751 board might affect the light yield determination in the chamber. 

\begin{figure}[ht]
 \centering
\includegraphics[width=9cm,height=6cm]{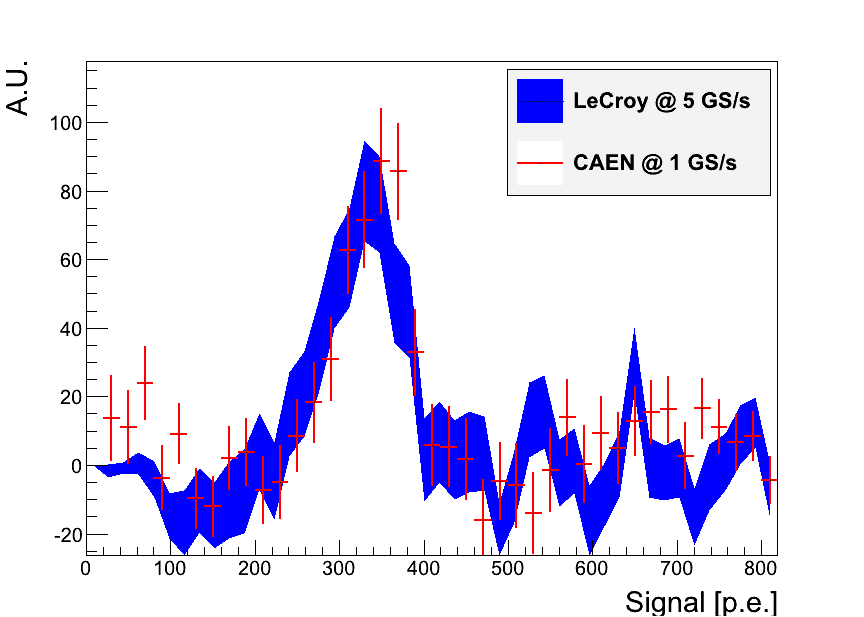}
 \caption{LY obtained using the LeCroy oscilloscope at 5~GS/s [ blue - band ] and the
V1751 board at 1~GS/s [red - points]. }
\label{fig:LY}
\end{figure}

For this test the waveforms acquired with the oscilloscope were of 8~$\mu s$ length with 1~$\mu s$
of pretrigger and 7~$\mu s$ of signal (about 5 times the
triplet decay length). Due to limited disk space, it was not possible to acquire waveforms with a sampling higher than 5~GS/s. 
 For each oscilloscope and V1751 run a SER spectrum was obtained in order to obtain the calibration factor
in $ADC\cdot ns/p.e.$ or $V\cdot ns/p.e.$ for the V1751 board data and oscilloscope data,
respectively. Once the calibration factor was known the total integral was
calculated by summing only integrals of peaks above a treshold of
three RMS of the
baseline region to discard noise peaks. For the selected peaks a
local baseline was subtracted 50~$ns$ before and after the signal. If in these
windows of 50 ns another peak was found it was merged with
the peak before it and the local baseline was recalculated. The total integral
in $ADC\cdot ns$ 
or $V\cdot ns$ was then normalized to the reference SER in order to obtain the
resulting spectrum in p.e.. The results of this operation for the V1751 board
and LeCroy oscilloscope data is presented in figure \ref{fig:LY}. The obtained
values were found to be in good agreement with each other about 3\%
difference in LY - well inside of measurement error suggesting both the satisfactory operation of
the V1751 fast ADC board and the sufficiency of using 1~GS/s sampling when
acquiring data from a liquid argon scintillation detector.

\section{Conclusions}

The V1751 is a fast ADC board manufactured by CAEN SpA. We have tested it in liquid argon Dark Matter detector
working conditions. We have found that the board 
is well suited for applications involving readout from a detector involving scintillation light registered by
fast PMTs. We have expanded the tests to determine whether the 1~GS/s and 2~GS/s sampling capability of
the board does not result in information loss in real data by comparing it to
a LeCroy WavePro 735Zi oscilloscope capable of up to 20~GS/s sampling for four
channels simultaneously. It has
been found that 1~GS/s sampling is adequate to record single
photolectron signals from cryogenic PMTs in liquid argon and that the bandwidth filter of the V1751 board is able to cut out most noise while maintaining the main signal features.

\end{document}